\documentclass[10pt]{article}
\pdfoutput=1 

\usepackage[round]{natbib}
\usepackage{subcaption}
\usepackage{float}
\usepackage{graphicx} 
\usepackage{url}

\usepackage{breakurl}
\usepackage[breaklinks]{hyperref}
\usepackage{amstext}
\usepackage{tabu}
\usepackage{multirow}
\usepackage{moreverb}
\immediate\write18{texcount -inc -incbib 
-sum research_notes.tex > /tmp/wordcount.tex}

\usepackage{titlesec}
\titleformat{\section}
  {\normalfont\fontsize{12}{15}\bfseries}{\thesection}{1em}{}
\titleformat{\subsection}
  {\normalfont\fontsize{11}{15}\bfseries}{\thesubsection}{1em}{}  

\usepackage[utf8]{inputenc} 
\usepackage[T1]{fontenc}    
\usepackage{booktabs}       
\usepackage{amsfonts}       
\usepackage{nicefrac}       
\usepackage{microtype}      
\usepackage{lipsum}

\title{Airbnb, hotels, and saturation of the food industry: A multi-scale GWR approach}

\author
{Zahratu Shabrina,$^{1}$ Boyana Buyuklieva,$^{1}$ Matthew Ng Kok Ming $^{1}$\\
\\
\normalsize{$^{1}$Centre for Advanced Spatial Analysis (CASA), University College London, UK}\\
\normalsize{$^\ast$To whom correspondence should be addressed; E-mail:  zahratu.shabrina.15@ucl.ac.uk.}\\
}

\begin{document}
\maketitle
\begin{abstract}
\noindent This paper evaluates the relationship between the food industry and the local variations of temporary accommodation (TAs, including hotels and short-term rentals). The aim is to capture the variance of the local statistic and pinpoint areas where food and beverages (F\&B) presence is highly related to TAs in London. We explain the phenomena using OLS and compare the result with the local model - Geographically Weighted Regression (GWR) and multi-scale GWR \citep{fotheringham2017multiscale} allowing the use of different optimal bandwidths instead of assuming that relationship varies at the same spatial scale. The comparison is presented and the result shows that the GWR model shows significant improvement over Ordinary Least Square (OLS), increasing the R\textsuperscript{2} from 0.28 to 0.75. MGWR further improves the model estimate, increasing the R\textsuperscript{2} to 0.77, indicating the relationship happens in different spatial scales. Lastly, as an estimate for F\&B, hotels appear to perform better in a high concentration of commercial and transport links functions, whilst Airbnb seems to perform better in highly residential areas proximate to the mainstream tourist attractions. Overall, this paper describes the use of the MGWR method in cases where localities is an important aspect of the spatial analysis process.

\noindent \textbf{Keyword:} GWR, MGWR, temporary accommodation, Airbnb, hotels, F\&B
\end{abstract}
\section{Introduction}

Simple linear regression, the most used technique in geographical analysis, assumes changes across space to be universal, which is not always the case in the spatial context. Variations across geographical space, known as spatial nonstationarity, might be lost when analysing using simple global fitting exercises such as Ordinary Least Squared (OLS) \citep{brunsdon1996geographically}. Therefore, geographically weighted regression (GWR), provides an alternative method to analyse and model the complex spatial variations in local parameter estimates \citep{fotheringham1997trends, brunsdon1996geographically, fotheringham1998geographically}. The developed method extends the traditional linear regression technique to incorporate spatial heterogeneity in different regions by allowing the parameter estimate to vary locally \citep{fotheringham2003geographically}. GWR has been used extensively to describe relationship in various fields, including but not limited to crime analysis \citep{cahill2007using}, population research (e.g drug resistance distribution \citep{shoff2012spatially}), epidemiology (e.g nutritional epidemiology \citep{yoo2012height} and infectious disease epidemiology \citep{liu2011spatial}), physical environment (e.g rainfall and altitude study \citep{brunsdon2001spatial}, land use and water quality study \citep{tu2011spatially}), and various other fields of study.

However, GWR has several limitations including the issue of multiple hypothesis testing that can give us results of false positives \citep{da2016multiple} and accuracy problem due to the assumption on a univariate scale by using a single bandwidth \citep{yu2019inference}. To overcome some of these issues, \cite{fotheringham2017multiscale}, \cite{yu2019inference} and \cite{oshan2018mgwr} proposed an extension of GWR method to include the computation of optimum bandwidths in each iteration of the local parameters, using multi-scale geographic weighted regression (MGWR). This new method improves the common GWR model mainly by eliminating the assumptions that variations occur within the same scale \citep{yu2019inference}. In contrast, MGWR allows multi-scale modelling, overcoming the issue of multiple testing as well as increasing the reliability by introducing multiple bandwidths \citep{fotheringham2017multiscale}.

This paper implements three models - OLS, GWR and MGWR - to analyse the relationship between competing temporary accommodation (TA) types in urban tourism - the traditional forms (hotels) and short-term rentals (Airbnb). The purpose is to explain how these establishments can be used as a proxy for the distribution of the food industry in various spatial locations in London. It adds our comprehension of the effect of spatial heterogeneity on industry location as local land-use within a city can be thought of as reactive to the impacts of urban tourism \citep{page1995urban}. As indicated in previous studies, there is a strong positive relationship between traditional accommodation, hotels, and places of interest in many cities \citep{lee2018spatial}. However, the rapid rise of short-term rentals (STR), through Online Platform Economy (OPE) such as Airbnb has somewhat changed the nature of tourism, especially in urban settings. Given that Airbnb can fundamentally be located wherever residential properties are available, it has been argued by \cite{guttentag2015airbnb} that this leads to increased dispersal of tourists in areas not typically regarded as central tourist destinations. \cite{zervas2016rise} expand on this, suggesting that Airbnb also pervade traditional tourist accommodation, such as hotels. However, the knowledge surrounding the impact on hotels and Airbnb to the retail land-use is limited; but, remains necessary, due to the significant contribution tourism provides on the city's highly competitive retail segment. Understanding the dynamics here might help us uncover interesting insight into the effects that temporary accommodation plays in changing its urban realm. Specifically for Airbnb, this call for an investigation is further strengthened given the fact London contains the greatest Airbnb supply globally (based on data from insideairbnb.com).

In our study, the hypothesis is that the interactions between TA and the food industry are not universal across space. Therefore we test our data using three different models: OLS, GWR and MGWR to describe the spatial distribution of Airbnb and hotels as determinants of the food industry in London. Previous studies have highlighted how hotels benefit strongly from the locational attributes of their proximate tourist and retail amenities \citep{arbel1977some,shoval2011hotel,shoval2004categorization}. Thus, in this paper, we investigate whether the local retail landscapes equally exhibit spatially-dependent distributions in relation to the existence of Airbnb. Following our study rationale, the remainder of our paper is organised as follows: firstly, the theoretical background of the relationship between Airbnb, hotels and F\&B point of interest is described; secondly, we evaluate these relationships using OLS, GWR, and MGWR respectively. We continue by presenting our result and discuss its implications on retail land-use. Lastly, we conclude the paper and provide a recommendation for further study.

\section{Inter-dependencies between Airbnb, hotels and food industry in urban tourism}

The Online Platform Economy (OPE) is a business model that monetises access to assets (i.e. spare rooms, tools, vehicles) which has been receiving increased attention over the past decade \citep{albinsson2012alternative,botsman2010s,agyeman2013sharing, guttentag2015airbnb}. Disruption by OPE is often viewed through the disruptive innovation lens (see \cite{christensen2015disruptive}, \cite{schmidt2008disruptive}, and \cite{gobble2016defining}) through the provision of alternative services to those already established within the industry. In the hospitality sector, Airbnb is a prominent OPE that offers rentals in a marketplace that connects hosts (with entire apartments, private or shared listings) and guests, with the added benefit of competitive pricing for potential guests. The impacts of this novel disruption on the hospitality industry, such as from short-term rentals (STR) have been widely documented. They are argued to serve the same function as hotels for which there is apparent overlapping demand within their targeted demographic \citep{CBRE2016, zervas2016rise}. \cite{zervas2016rise} found an increase in Airbnb supply came with an estimated decline to hotel revenues of up to 8-10\%. This has been particularly prominent within lower-end categories of hotels that may not cater to the needs of the entire cross-section of potential tourists and travellers \citep{zervas2016rise}. This competitive impact of Airbnb over traditional hotels was again substantiated in a recent study by \cite{dogru2017hotel}, which indicated a revenue decline of approximately 2.5\% across all hotels in Boston.

\cite{arbel1977some} found that the travel patterns of tourists within a city remain close to the vicinity of their chosen accommodation and their intent to visit individual areas of interest. This is a result that has been expanded upon in research by \cite{shoval2011hotel}, illustrating that large shares of tourist travel within a city often lies in a definable area that extends over their chosen accommodation. It can, therefore, be assumed that areas with high-clusters of hotels are often faced with the pressure to accommodate the increased temporal influx of tourist populations. This is most evident in the oversupplied and saturated city facilities (transport links, entertainment centre, etc.) within many central tourist areas \citep{shoval2004categorization}. The intensified use of these areas by tourists suggests there is increased retail opportunity \citep{yang2014theoretical}; which corresponded to changes in business and retail land-use, like an increased number of F\&B establishments, to meet the increased tourist demands. These studies call for a more in-depth review of policy mechanisms that deal with the integration of increased tourists movement with that of the land-use attributes of localised areas, and this paper serves to fill this gap.

Although findings from previous studies have pointed to a strong spatial relationship between the distribution of hotels and constructed tourist amenities, indicating the critical role tourist amenities play within proximate locations to hotels \citep{lee2018spatial, li2015spatial}, this relationship is not very clear for STRs such as Airbnb. Several studies certainly suggested that the presence and availability of the Airbnb platform have steadily fulfilled a growing need for alternative accommodation within the hospitality industry \citep{guttentag2018tourists, zervas2016rise}. As such, the overlap of potential customers for all forms of temporary accommodation (TAs), whether provided by OPE platforms or hotel, is said to have a significant increase over recent years \citep{dogru2017hotel,gutierrez2017eruption}. This is evident in both in the rising trend of hotels offering typical Airbnb attributes through purchase and partnerships \citep{solon_2018}; and, similarly, with Airbnb following this trend to capture a wider market share through the acquisition of other hotel booking platforms \citep{ting_2018}. However, despite the growing supply of TAs through these forms of consolidation, the documented relationships and implications of Airbnb on local land-use remains limited; particularly, in comparison to the extensive literature on hotel spatial distributions. More specifically, whilst the spatial pattern of hotels can often be predicted by several factors, such as local zoning and planning regulations that further limit their densities and distribution, the impacts of new forms of accommodation types like those arising from STRs are largely still unknown. The primary concern raised in this paper is that of the relationship between STRs and the traditional hotel accommodation in the context of understanding current local land-use. In particular, we look at retail land-use through the lens of available food and beverage (F\&B) establishments as an indicator for this relationship as they serve to cater both tourists and residents alike. 

\section{Overview of the dataset used}
We use three dataset to describe the relationship of TAs, measured in Airbnbs and hotels, with F\&B establishments. The first dataset is the traditional accommodation data (including guest houses, bed and breakfast, hostels, hotels, motels, country houses, inns, youth hostels, and other youth classifications) from the Ordinance Survey Points of Interest (POI) data. From the same data source, we also use data on F\&B establishments, including restaurants, fast food outlets, pubs, and other similar types of venues. Our last dataset contains Airbnb listings, as a form of STR using the online platform, from Inside Airbnb (http://insideairbnb.com). These include private rooms, shared rooms, and entire homes.

\begin{figure}[!ht]
    \centering
    \begin{subfigure}{0.32\textwidth}
            \includegraphics[width = 1\textwidth]{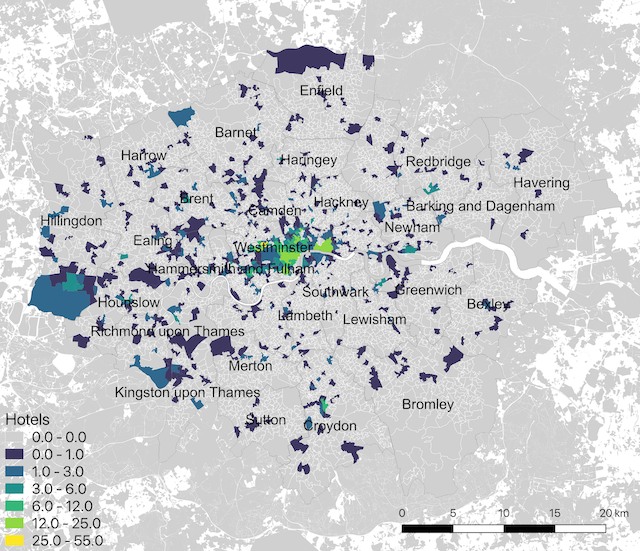}
            \caption{Hotels are concentrated in multiple areas}
            \label{fig:hotels}
    \end{subfigure}
    \begin{subfigure}{0.32\textwidth}
        \centering
            \includegraphics[width = 1\textwidth]{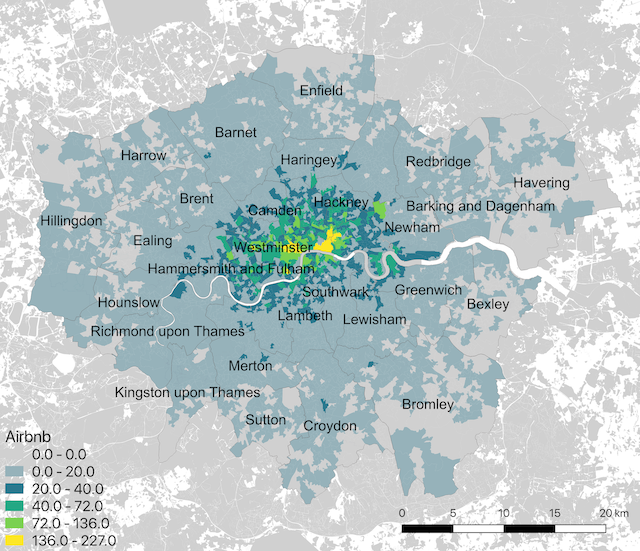}
            \caption{Airbnb are concentrated in central areas}
            \label{fig:airbnb}
    \end{subfigure}
    \begin{subfigure}{0.32\textwidth}
        \centering
            \includegraphics[width = 1\textwidth]{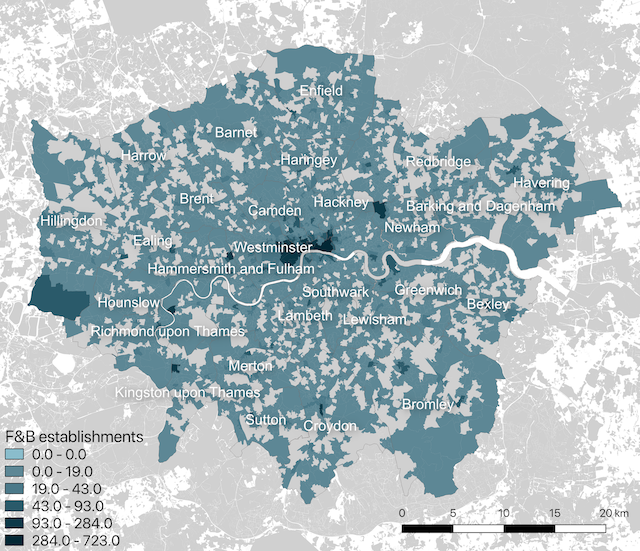}
            \caption{F\&B are geographically dispersed}
            \label{fig:fb}
    \end{subfigure}
    
\caption{The spatial distribution of the temporary accommodations (predictor variables (a,b) and outcome variable F\&B (c)).}
\label{fig:distribution}
\end{figure}

Figure \ref{fig:distribution} shows the data across London. It illustrates the concentration of our study's 1382 hotels distributed in only 644 LSOAs (13\% of the total LSOA) in Figure \ref{fig:hotels}. Less extreme is the distribution of Airbnb. The data contains over 69,000 Airbnb listings; however, we only consider active listings (these are proxied by the presence of at least one review). Airbnb which satisfy this condition is aggregated for each LSOA in London for a total of 50,000 listings across 3982 LSOAs (82\% of all LSOAs) (Figure \ref{fig:airbnb}). Lastly, Figure \ref{fig:fb} shows the spatial distribution of 27,716 F\&B establishments in London, distributed in 3306 LSOA (68\% of the total LSOA). A summary of these counts is provided in Table \ref{tab:data_points}. Due to a different representation of data, for example, an Airbnb can be shared, private or entire home listing with small capacity for occupants, while a hotel might actually represent hundreds of room, we use z-score to standardise our data value. We re-calculate the data distribution to have a mean of 0 and a standard deviation of 1, as proposed by various studies using GWR method \citep{fotheringham2017multiscale, oshan2018mgwr, yu2019inference}. This method is proven more effective than other standardisation methods generally \citep{milligan1988study}. What it does is it recalculates each value and based on the data mean and standard deviation, so then the data is redistributed according to below average, near average, and above average. 

\begin{table}[!ht]
  \small
  \centering
    \begin{tabular}{l|cc}
    \textbf{Type}  & \multicolumn{1}{l}{\textbf{Total Count}} & \multicolumn{1}{l}{\textbf{LSOA Containing (\%) }} \\
    \midrule
    Airbnb &           50 000  & 82\% \\
    Hotel &              1 382  & 13\% \\
    F\&B &           27 716  & 68\% \\
    \end{tabular}%
  \caption{Overview of the data for the study area. The latter is all 4835 LSOAs in central and greater London.}
  \label{tab:data_points}%
\end{table}%

\subsection{Statistical base-lining}
Table \ref{tab:multicollinearity} shows the descriptive statistic of the data as well as the variance inflation factor (VIF). VIF assesses how much variances increases if predictors are correlated with no correlation would yield VIF score of 1 and a VIF between 5 and 10 indicates problematic multicollinearity \citep{wheeler2005multicollinearity, wheeler2007diagnostic}. From Table \ref{tab:multicollinearity} we can see that both covariates are not exhibiting multicollinearity with VIF close to 1. 
\begin{table}[ht]
\small
  \centering
    \begin{tabular}{l|ccccc}
\textbf{Variable} & \textbf{Mean}   & \textbf{SD}    & \textbf{Min} & \textbf{Max} & \textbf{VIF}  \\
\midrule
Airbnb  & 10.34  & 17.30  & 0   & 227 & 1.19 \\
Hotel    & 0.29 & 1.45  & 0   & 55  & 1.19 \\
F\&B     & 5.73  & 16.36 & 0   & 723 & -    \\
\end{tabular}
\caption{Descriptive statistics.}
  \label{tab:multicollinearity}%
\end{table}

We set a baseline of the relationship between TAs and F\&Bs using a multivariate linear regression with two independent variables, namely Airbnb and hotels. The result of this regression is reported in Table \ref{tab:regression} suggesting a positive statistically significant relationship between the presence of F\&B establishments and both TA types. However, the model exhibits poor performance, explaining only 28\% of the variation in the data. As a next step, we exclude significant outliers, which are defined as those beyond a far outer Tukey Fence (i.e. 3 times the interquartile range). Such outliers are present in the hotel and F\&B data: Westminster, where there is an area with a total of 55 hotels; and, in the City of London, with 723 F\&B establishments.


\begin{table}[htbp]
  \centering

    \begin{tabular}{lll}
    & \textbf{lm - with outliers} & \textbf{lm - without outliers}\\
    \midrule
    Intercept & 1.135*** & 1.622*** \\
   Hotel & 2.578*** &  2.545*** \\
    Airbnb & 0.373*** & 0.317*** \\
    \textbf{R\textsuperscript{2}} & \textbf{0.281} & \textbf{0.327}\\
    \multicolumn{3}{l}{\textit{Significance level: *** p-value \textless{} 0.05}} \\
    \multicolumn{3}{l}{\textit{Hotel outlier: Westminster}} \\
    \multicolumn{3}{l}{\textit{F\&B outlier: City of London}} \\
    \end{tabular}%
    \caption{The global model using complete data (with outliers) and without outliers.}
  \label{tab:regression}%
\end{table}%

\begin{figure}[!ht]
\centering
    \includegraphics[width=0.6\textwidth]{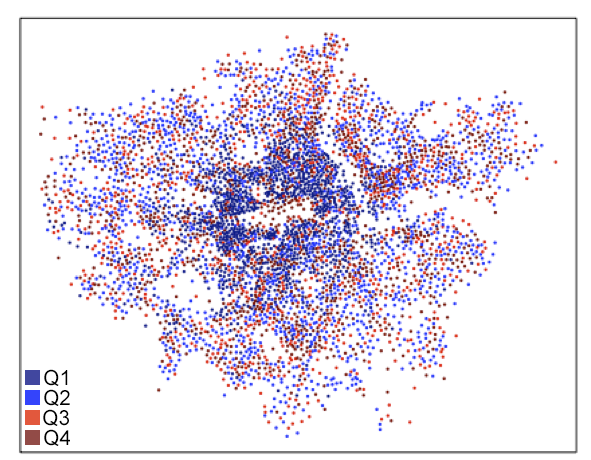}
    \caption{Spatial Residuals plotted by quantiles.}
    \label{fig:resids}
\end{figure}

We then check our data for spatial autocorrelation using Moran's I. Spatial autocorrelation captures the spatial relationship of a dataset, where closer observations are more related than distant ones \citep{anselin2001spatial}. Our result for Moran's I test indicates that the residuals show spatial dependence (p-value of $<$ 0.01). We can also see the effect of spatial auto-correlation in Figure \ref{fig:resids}. There is a clear pattern in space: highest under predictions (dark blue) are clustered in central London, whereas over predicted residuals (red) are in the periphery, with smaller clusters in the northern and southwestern boroughs. Based on the above and to create a more robust estimate, we disregard the global regression model. The problem with this approach is that heterogeneity is lost across space. Global statistics, such as the multiple linear regression used above, generalise across a study area, reducing the inherent variability at localised scales. The addition of a spatial facet diminishes 'spatial stationarity' \citep[p.89]{fotheringham1997trends} in the data; and, therefore, our methodology needs to be adapted to account for this. Geographically weighted regression (GWR) is a method to localise regression modelling to understand the relationships between variables across space, thereby addressing the issue of spatial autocorrelation \citep{brunsdon1996geographically, fotheringham1997trends, fotheringham1998geographically}.

\section{Analysing spatial heterogeneity using GWR and MGWR methods}
GWR extends traditional regression analysis to allow local (instead of global) parameters to be estimated. Given a dependent variable $y$ in location $i$ where $u_{i},v_{i}$ denotes the coordinates in point $i$ with continuous function $a_{k}(u,v)$ at point $i$ \citep{brunsdon1996geographically, fotheringham1997trends, fotheringham2003geographically}, GWR model can be written as follow (Equation \ref{eq:gwr}):

\begin{equation}
y_{i} = a_{0}(u_{i},v_{i}) + \sum_{k}^{} a_{k}(u_{i},v_{i}) x_{ik} + \epsilon_{i}
\label{eq:gwr}
\end{equation}

We conduct GWR using spgwr package within the R ecosystem (for complete methodology check \cite{bivand2017package}). In analysing our data using GWR method, we first need an understanding of what is local. We do this by specifying bandwidth, for which to take data points into account for the local regression model. The GWR measure the observation in accordance with the proximity to point $i$, which is LSOA in our case study. This bandwidth value is computed based on each LSOA point so as to find the optimum range for which to include other data points, whilst minimising the error of local model at each point. Therefore, for many possible bandwidth values,  $\beta$ , the procedure minimises the function  $s ( \beta )$, which is the prediction error model at each point as shown in Equation \ref{eq:err}.

\begin{equation}
s ( \beta ) = \sum _ { i = 1 } ^ { n } \left\{ y _ { i } - \hat { y } _ { \neq i } ( \beta ) \right\} ^ { 2 }
\label{eq:err}
\end{equation}

Then, we determine the weights of for each observation $w _ { i j }$, relative to all regressions at the F\&B points. As our data contains outliers and long-tailed error distribution, we use an adaptive bi-square kernel (refer to \cite{charlton2009geographically}, p5-8), where only observations inside the bandwidth are taken into account, nullifying the rest. An adaptive kernel is more favourable when dealing with non-uniform spatial distributions \citep{oshan2018mgwr}. The weight of data point $j$ at regression point $i$ ($w _ { i j })$ based on the distance between regression point $i$ and data point $j$ ($d _ { i j }$) can be calculated as follow in Equation \ref{eq:distance}. The optimal bandwidth size is calculated based on the number of neighbouring observations and the optimal proportion is returned by minimising the Akaike Information Criterion (AICc) score \citep{fotheringham1998geographically}.

\begin{equation}
w _ { i j } = {[1 - ((d _ { ij } / b) ^ { 2 } ]^2} \\\
if \\\ d _ { ij } < b , 0 \\\ otherwise
\label{eq:distance}
\end{equation}

Because GWR assumes bandwidth is constant for each relationship, there are possibilities for misspecification of one or more scales \citep{oshan2018mgwr}. Thus, MGWR further extends the functionality of GWR by allowing the use of specific bandwidth for each variable denoted as $ bw $ allowing the model to operate at different spatial scales \citep{fotheringham2017multiscale}. Using back fitting algorithm, an iterative procedure to fit generalised additive model (GAM), MGWR relating a univariate response variable (in this case the GWR model) to predictor variable (partial residuals from previous iteration), using smooth function until the model converges to a solution \citep{fotheringham2017multiscale, yu2019inference}. We use a Python based package called mgwr (for complete methodology check \cite{oshan2018mgwr}). The computation is based on iterations of optimum bandwidth for each parameter estimates. MGWR model can be written as follow (Equation \ref{eq:mgwr}):

\begin{equation}
y_{i} = \beta_{bw0}(u_{i},v_{i}) + \sum_{k}^{} \beta_{bwk}(u_{i},v_{i}) x_{ik} + \epsilon_{i}
\label{eq:mgwr}
\end{equation}

\section{Results and discussion} 
As the previous OLS model exhibits poor performance and the residuals indicate spatial autocorrelation, it is not an ideal model to analyse the relationship of Airbnb, hotels and F\&B locations. Therefore, this section focuses on the result of the later models used in explaining the spatial heterogeneity through geographically weighted regression (GWR) and multi-scale geographically weighted regression (MGWR).

\subsection{Model comparison and performance}
We conduct two different models: GWR and MGWR.  Table \ref{tab:result} shows the comparison between the result of the implemented models. It showcases the descriptive statistic of the parameter estimates and the model's goodness of fit. Based on the analysis, MGWR performs slightly better, as shown by lower Corrected Akaike Information Criterion (AICc), lower Residual Sum of Square (RSS), and higher R\textsuperscript{2}, compared to GWR. Generally, AICc is used for model selection in GWR \citep{charlton2009geographically}, representing the relative amount of lost information in the model by considering both the risk of overfitting and underfitting \citep{akaike1998information}. The AICc for MGWR is 7656, compared to 8077 for GWR, showing that the MGWR model is of higher quality due to the lower level of model losses. The RSS for MGWR is also lower, meaning that the model has less discrepancy between the observed and estimated values. The R\textsuperscript{2} improved slightly when changing from GWR to MGWR, thus implementing different bandwidth for each variable allows the model to explain more variance in the model. Table \ref{tab:result} showcases that a single bandwidth of 69 is used for the GWR model, while the MGWR bandwidths show different spatial scales - 73 for Airbnb and 43 for hotels based on the calculation after 30 iterations. The performance of both models depends largely on the optimum use of bandwidths or neighbouring values for model smoothing where each local estimation is based \citep{charlton2009geographically}. By allowing multiple bandwidths in MGWR, the model accounts for an optimal number of neighbours for each parameter estimate, providing more accurate predictions for the response variables.

\begin{table}[ht!]
\centering
\begin{tabular}{lllllllcccc}
\toprule
\multicolumn{1}{l}{\textbf{Model}} & \textbf{Variable} & \textbf{Mean} & \textbf{STD} & \textbf{Min} & \textbf{Median} & \textbf{Max} & \textbf{Bandwidth}  & \textbf{R2}           & \textbf{RSS}              & \textbf{AICc}             \\
\midrule
\multirow{3}{*}{\textbf{GWR}}      & Intercept         & 0.085         & 0.423        & -3.491       & -0.002          & 2.421        & \multirow{3}{*}{69} & \multirow{3}{*}{0.75} & \multirow{3}{*}{1199}  & \multirow{3}{*}{8077}   \\
                                   & Airbnb            & 0.513         & 0.722        & -1.596       & 0.323           & 4.454        &                     &                       &                           &                           \\
                                   & Hotel             & 0.318         & 0.586        & -14.541      & 0.205           & 4.222        &                     &                       &                           &                           \\
\midrule
\multirow{3}{*}{\textbf{MGWR}}     & Intercept         & -0.045        & 0.053        & -0.166       & -0.043          & 0.04         & 1392                & \multirow{3}{*}{0.77} & \multirow{3}{*}{1133} & \multirow{3}{*}{7656} \\
                                   & Airbnb            & 0.252         & 0.218        & -0.433       & 0.226           & 1.411        & 73                  &                       &                           &                           \\
                                   & Hotel             & 0.277         & 0.451        & -1.277       & 0.206           & 3.94         & 43                  &                       &                           &                          \\
\bottomrule                                   
\end{tabular}
\caption{GWR and MGWR result for spatial variability of parameters.}
\label{tab:result}%
\end{table}%

\begin{figure}[!ht]
    \centering
    \begin{subfigure}{0.48\textwidth}
        \centering
        \includegraphics[width=1\textwidth]{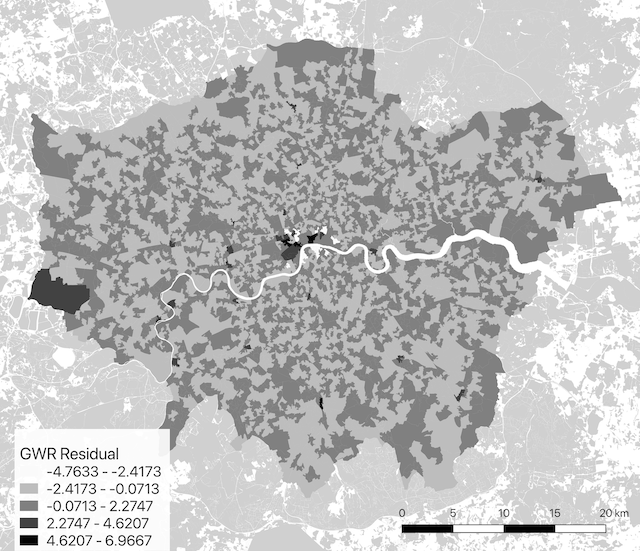}
        \caption{Non-random pattern of GWR residual}
        \label{fig:gwr_res}
    \end{subfigure}
    \begin{subfigure}{0.48\textwidth}
        \centering
        \includegraphics[width=1\textwidth]{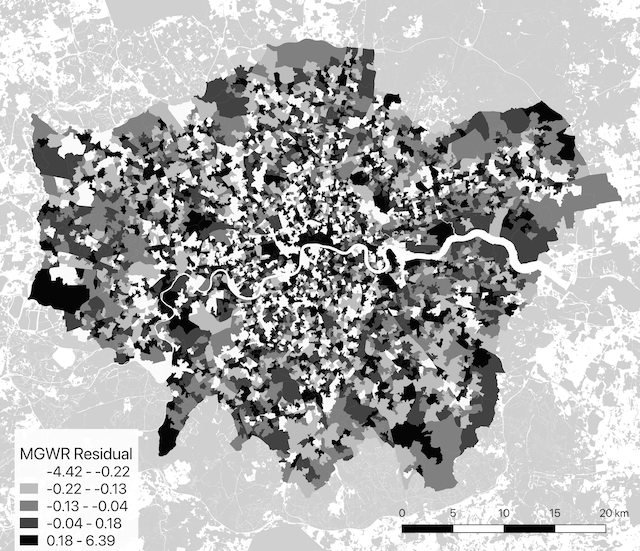}
        \caption{Random pattern of MGWR residual}
        \label{fig:MGWR_res}
    \end{subfigure}
\caption{Spatial distribution of GWR and MGWR residuals. The GWR residuals (left) still shows some clustering in central and west London, while MGWR residuals (right) are completely randomly distributed.}
\label{fig:residual}
\end{figure}

Next, we plot the residuals from both models in order to Figure \ref{fig:residual} shows the comparison between model residuals calculated from the difference between observed values and estimated values returned by both GWR and MGWR. We can observe that the GWR residuals still show some spatial clusterings, especially in inner London and west London near Heathrow airport, meaning that the GWR model is unable to explain the observation optimally. In contrast, MGWR residual appears to be distributed randomly, showcasing that only random errors are left. Thus, based on the previous analysis of the model performance, as well as the spatial distribution of the model residuals, MGWR is more favourable compared to GWR and the next section will focus on the result of MGWR model.

\subsection{MGWR model interpretation}
\begin{figure}[!ht]
    \centering
    \includegraphics[width=0.8\textwidth]{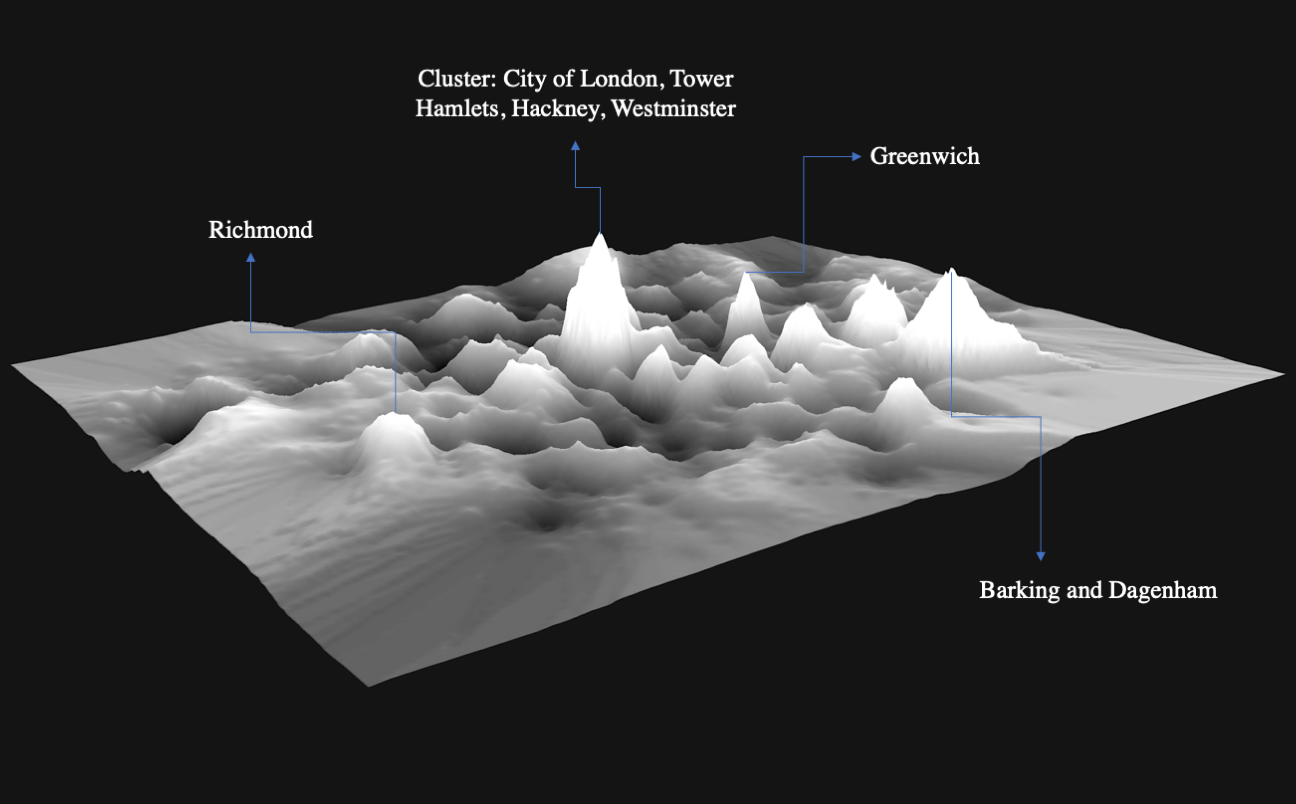}
    \caption{Elevation parameter surface of the MGWR estimates if we were to increase ecah type of TA by one unit showing local variations of the data.}
    \label{fig:3D}
\end{figure}

Figure \ref{fig:3D} presents the 3D visualisation of the MGWR parameter estimates based on our previous model below:
\begin{equation}
F\&B_{mgwr} = ( -0.166 \: to \: 0.04 ) + (-0.433 \: to \: 1.411)*A  + (-1.277 \: to \: 3.94)*H + \epsilon_{i}
    \label{eq:mgwr_result}
\end{equation}

The method to incorporate the MGWR parameter estimates (the range of minimum and maximum values) into the MGWR model shown in Equation \ref{eq:mgwr_result} is used previously by \cite{brunsdon2001spatial} to show how the estimates vary across space for each variable. The parameter surface in Figure \ref{fig:3D} represent the concentration of predicted F\&B establishments using the model parameters if we were to increase both TAs (Airbnb and hotels) by one unit in the MGWR model. We can see from the 3D visualisation, how the parameter estimates have spikes (higher values) and lower surfaces in other areas, indicating that the relationship of temporary accommodations towards F\&B is not constant across space, but instead they are nonstationary.

\begin{figure}[!ht]
    \centering
    \begin{subfigure}{0.48\textwidth}
        \centering
        \includegraphics[width=1\textwidth]{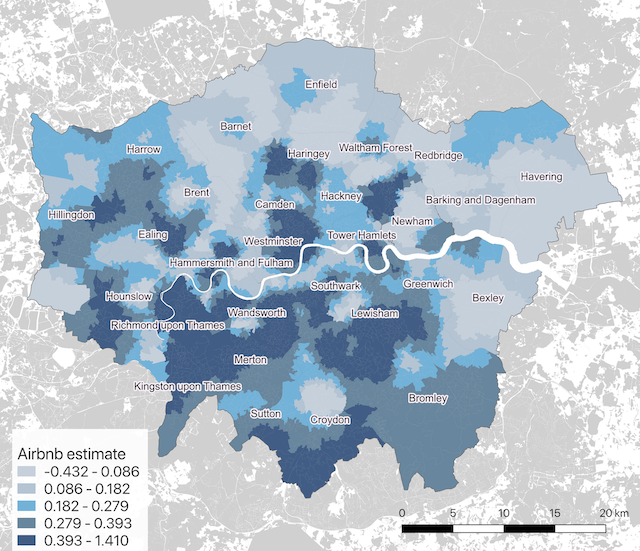}
        \caption{Airbnb estimate (MGWR bandwidth = 73)}
        \label{fig:airbnb-estimate}
    \end{subfigure}
    \begin{subfigure}{0.48\textwidth}
        \centering
        \includegraphics[width=1\textwidth]{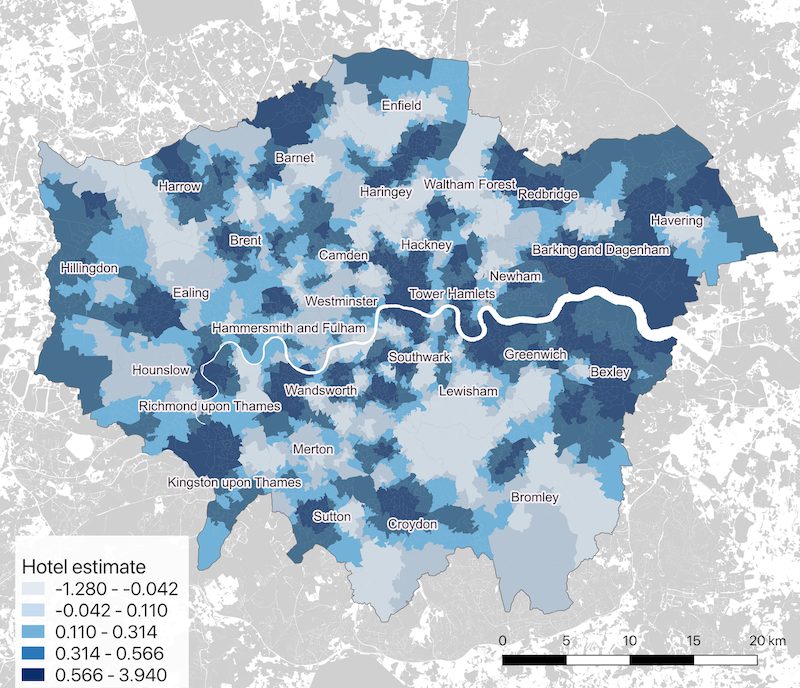}
        \caption{Hotel estimate (MGWR bandwidth = 43)}
        \label{fig:hotel-estimate}
    \end{subfigure}
    \begin{subfigure}{0.48\textwidth}
        \centering
        \includegraphics[width=1\textwidth]{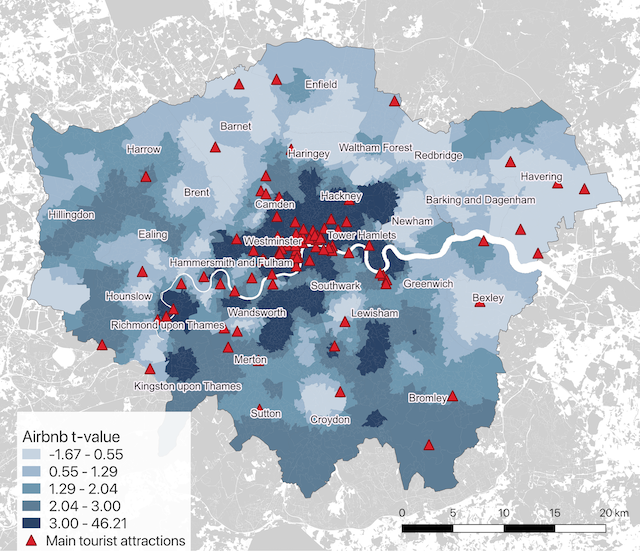}
        \caption{Airbnb t-value}
        \label{fig:airbnb-t}
    \end{subfigure}
    \begin{subfigure}{0.48\textwidth}
        \centering
        \includegraphics[width=1\textwidth]{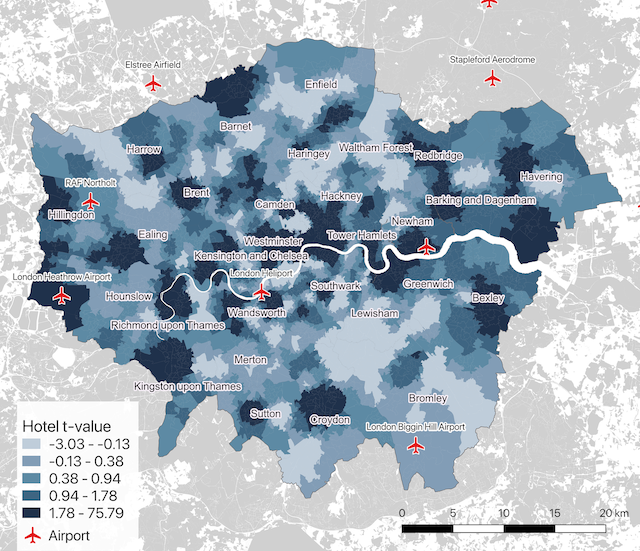}
        \caption{Hotel t-value}
        \label{fig:hotel-t}
    \end{subfigure}
\caption{The MGWR model results showing the parameter estimates and the level of significance as indicated by the distribution of t-values. The model showcases how Airbnb and hotels have distinct relationships with F\&B locally in different spatial scales.}
\label{fig:t-value}
\end{figure}

To gain a more detailed understanding of the local variation, in this section, we present the result of the model. The MGWR model shows that the relationship between Airbnb and F\&B establishments happens at a larger spatial scale compared to hotels. Figure \ref{fig:airbnb-estimate} and \ref{fig:hotel-estimate} shows the parameter estimates of the variables using the different bandwidth assigned by the MGWR model. All the spatial distribution in this paper is mapped using quantiles range. As MGWR provide us with an abundance of information, another measure beside parameter estimates is needed. Figure \ref{fig:t-value} also shows the level of significance for Airbnb (Figure \ref{fig:airbnb-t}) and hotel (Figure \ref{fig:hotel-t}), where the darker areas show higher t-values corresponding to higher significance. T-values increases along with an increase of difference between sample data and the null hypothesis.

From Figure \ref{fig:t-value}, we can see that Airbnb and hotels show relationship with F\&B not just at different spatial scales, but also different spatial distribution. For example, Airbnb shows a positive relationship to the areas in the southwest and some in south London, but the t-values are only high in central to southwest areas, excluding the positive estimate in south London. Airbnb in areas such as Westminster, Camden, Lewisham and Richmond upon Thames can be used to indicate the high the amount of F\&B in those areas.  Both the positive relationship and the larger t-values are located proximate to the main tourist attractions (shown as red triangles in Figure \ref{fig:airbnb-t} based on the data from Visit Great Britain, 2015). Spanning from central to southwest London are the locations of London main tourist attractions including British Museum, Kensington Palace, Natural History Museum, Kew Garden, and Richmond Park.

By contrast, t-values for hotels are generally more dispersed. Areas, where traditional hotels prevail, are not just in areas with larger venues/attractions, but also ones with very specific transport links. The obvious example is the higher t-values in the area around Heathrow Airport, including Hillingdon and Ealing as the main entrance of air transport to London. The smaller patch in Hillingdon is also in comfortable walking distance to the Royal Air Force base and the Northolt Jet Centre (an extension of London City Airport). Similarly, the area between Brent and Barnet has a quick link to Luton Airport (direct 30 minutes transfers to the airport from Hendon Station - (Cited travel times are from Google Maps, for a weekday in 2019)). The blue patches in Redbridge and Barking and Dagenham are centred around Dagenham Heathway and Chadwell Heath Stations, respectively which have rail links to Southend Airport. Areas adjacent to London City Airport also shows higher t-values, such as Newham and Greenwich. There is evidence for the case that Hotels and F\&Bs co-vary in areas that cater to a specific niche of air travellers. Moreover, the MGWR result also shows higher significance in areas along the Thames River.

\begin{figure}[!ht]
    \centering
    \begin{subfigure}{0.48\textwidth}
        \centering
        \includegraphics[width=1\textwidth]{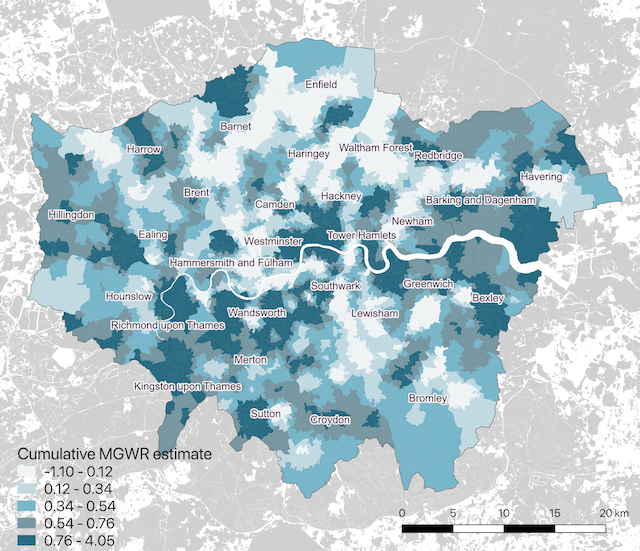}
        \caption{Cumulative estimate when both Airbnb and hotels are increased by one unit}
        \label{fig:cumulative_coef}
    \end{subfigure}
    \begin{subfigure}{0.48\textwidth}
        \centering
        \includegraphics[width=1\textwidth]{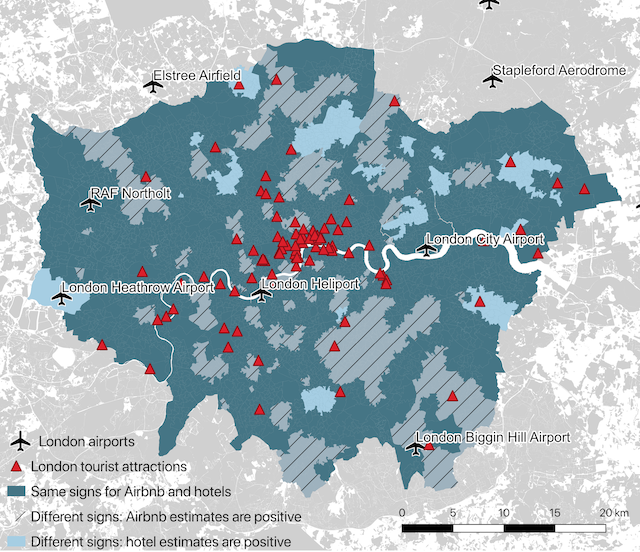}
        \caption{Different signs of parameter estimates proximate to airports and tourist attractions}
        \label{fig:diff_sign}
    \end{subfigure}
\caption{Detailed examination of GWR result based on cumulative estimates and different sign parameter estimates.}
\label{fig:GWRresult}
\end{figure}

Figure \ref{fig:cumulative_coef} shows cumulative parameter estimates to assess the relationship of F\&B with TAs simultaneously. Among the areas with positive cumulative estimates are the clustering of City of London, Hackney, Tower Hamlet and Westminster. These are the main areas in London with concentration of housing, commerce and tourist attractions. Greenwich, an area rich in cultural history that offers mix of urban tourism (such as Greenwich Park, multipurpose O2 arena for sports, concert, etc), offices and residential area, also shows positive relationship between TAs and F\&B. The MGWR can pick up these localities within the model.

Another observation is that TAs do not always show the same relationship to F\&Bs. Figure \ref{fig:diff_sign} shows areas where the sign of the parameter estimate for hotels differed from Airbnb. These are areas where Airbnb and hotels have different relationships with F\&B, in contrast with one another. The areas with diagonal pattern are those with positive Airbnb estimates and a negative coefficient for hotels. These are mostly in patches of residential areas north and south London, where most of the areas are not serviced by hotels, i.e. Lewisham and Enfield. But they are also areas with some larger attractions that cater to visitors such as Westminster and Camden.

Conversely, we find that LSOA for which hotel estimates are positive and Airbnb estimates are negative have close links to air travel facilities either through simple geographic proximity or convenient transport links. Hotels appear to be prevalent in areas close to airports. It could be that hotels take precedent over Airbnb in these areas because of the former offer more expansive support (e.g. flexible check-in hours, larger facilities and other services), which would be convenient to air travellers. This claim could be supported by further work that more specifically examines what services hotel offer that would be of value to air travellers.

There are two limitations to our study. Firstly, we consider a fairly limited number of variables, making our GWR analysis relatively focused. Whilst the model shows a strong relationship between these variables, it may also be prudent to include other sub-sectors within the retail industry to improve the study results. Also, as previously suggested, the study can be augmented using additional data in particular demographic variables. Secondly, we only analyse a single snapshot of the phenomenon thus it is not possible to present any changes in land-use associated with the presence of TAs. Based on this, a suggestion for further work would be to consider this analysis using time-series data when this is available.

Despite the limitations mentioned above, our study contributes to the development of current urban studies especially related to tourism and land-use. This paper serves as an example of the modelling of the relationship of TAs (Airbnb and hotels) in relation to F\&B in London, as an example of shared city land-use between visitors and residents. The GWR and MGWR results raise the point that the interaction between TAs and F\&Bs is complex and happened at localised levels.

\section{Conclusions}
In urban tourism, the impact of hotel presence in cities are widely studied for example in relationship with the distribution of local amenities e.g. food industry. Localities are important as location is a highly prominent aspect for hotel spatial locations. However, the rise of digital platforms in hospitality industry for the past decade has given a new nuance to the overall temporary accommodation sector, but this addition is largely understudied. That is why, our study offers a synoptic view of London in terms of the relationship between F\&B and TAs, including Airbnb (as an example of Internet based short term rental) and hotels (traditional tourist accommodation). Upon conducting three different models to measure this relationship, we find that spatial heterogeneity is lost in OLS, thus a more robust model is needed. The GWR and MGWR both improve the model performance, by allowing the local variations to be captured. The latter model, MGWR, further allows the use of different bandwidth to be used, so the analysis can be conducted in different spatial scales. This method can be implemented to analyse phenomena which variables vary locally and exhibit spatial non-stationarity in different levels.

Generally, as also the case of hotels, F\&B exhibits spatial dependence to Airbnb as indicated in our analysis. Thus, both Airbnb and hotels can be an indicator for saturation of the food industry in cities as they provide pressure to city facilities especially those that cater to the needs of both visitors and residents. Our MGWR model provides us with two main findings. Firstly, although both Airbnb and hotel location can provide an estimate in the saturation of the food industry, the two types of temporary accommodation exhibit relationships with F\&B across two different spatial scales. By allowing the use of different optimum scales among the two variables in our analysis, we are able to assess the complexity of the effect of both types of temporary accommodations on the food industry. The relationship between Airbnb and F\&B happens in a larger spatial scale, indicated by a larger bandwidth of 73. This means that for the case of Airbnb, the model needs a larger amount of neighbouring values for the model to perform optimally. This is possibly due to the spatial locations of Airbnb that is monocentric yet distributed all across the city with very high numbers. So, compared with the hotel locations that have multiple concentrations, but scarce overall (in terms of distribution, many areas in cities are not serviced by hotels), the use of smaller bandwidth provides a more optimum result. Thus, the relationship between hotel and F\&B can be analysed in a smaller spatial scale. 

Secondly, both types of temporary accommodations has different relationships with F\&B . While Airbnb have a stronger relationship with the local F\&B land use in areas with a predominantly residential built environment fabric near the tourist attractions, hotels have a stronger relationship with transport links such as airports. This suggests that, from a land-use perspective, the F\&B and possibly wider local retail in these areas might be more sensitive to visitor needs, rather than local ones. Therefore, this finding could justify the need to put extra care in issuing land-use permissions and short-term rental regulations in these areas to ensure a balance is struck between local and visitor needs.

To conclude, our study shows that we are not able to capture the complexities of the interactions between F\&B and temporary accommodations using simple regression as the overall relationship is not universal. Using GWR, the model specification improves significantly showing that local retail landscapes equally exhibit spatially-dependent distributions in relation to the existence of both hotels and Airbnb. But, MGWR model further shows that these interactions happen in different spatial scales for hotels and Airbnb where Airbnb shows the relationship with  F\&B at a larger spatial scale compared to hotels. The proliferation of online tourist accommodation in residential areas prove to subject these areas with similar pressure for local industries as hotels do. Th. Gis study shows that MGWR is performing well in capturing this purpose.

\section*{Acknowledgement}
This work was supported by Indonesia Endowment Fund for Education (LPDP) Ministry of Finance, Indonesia.
\bibliographystyle{apalike}
\setlength{\bibsep}{1pt plus 0.3ex}


\end{document}